\documentclass[12pt,a4paper]{article}
\usepackage[russian,english]{babel}
\usepackage[cp1251]{inputenc}
\usepackage[T2A]{fontenc}
\usepackage{latexsym}
\usepackage{amsmath}
\usepackage{graphicx}

\newcommand{\be}{\begin{equation}}
\newcommand{\ee}{\end{equation}}
\newcommand{\p}[1]{(\ref{#1})}
\def\a{\alpha}
\def\h\a{\hat\a}
\def\b{\beta}
\def\h\b{\hat{\b}}
\def\p{\partial}

\def\t{\theta}
\def\tr{{\rm tr}}
\def\tr{{\rm tr}\,}
\def\Tr{{\rm Tr}\,}
\def\cN{{\cal N}}
\def\cD{{\cal D}}
\def\cA{{\cal A}}

\def\bea{\begin{eqnarray}}
\def\eea{\end{eqnarray}}

\def\cN{{\cal N}}

\def\cF{{\cal F}}

\def\cW{{\cal W}}
\def\f{\frac}

\def\tr{{\rm tr}\,}

\def\bea{\begin{eqnarray}}
\def\eea{\end{eqnarray}}

\def\s{\sigma}

\def\d{\delta}
\def\q{\quad}
\def\g{\gamma}
\def\G{\Gamma}
\def\S{\Sigma}

\def\l{\label}
\def\ve{\varepsilon}

\def\sB{\stackrel{\frown}{\Box}}

\def\l{\ldots}
\def\cF{\cal F}

\date{\it  }
\begin{document}

\begin{center}
\vspace{1cm} {\Large\bf  Effective actions in $\cN=1$, D5
supersymmetric gauge theories: harmonic superspace approach
\vspace{1.2cm} }

\vspace{.2cm}
 {I.L. Buchbinder$^{a,}$$^{b}$,
 N.G. Pletnev$^{c,}$$^{d}$
}

\vskip 0.6cm { \em \vskip 0.08cm \vskip 0.08cm $^{a}$Department of
Theoretical Physics, Tomsk State Pedagogical University\\
Tomsk, 634061 Russia \\
$^{b}$National Research Tomsk State University;\\
joseph@tspu.edu.ru \vskip 0.08cm \vskip 0.08cm $^{c}$Department of
Theoretical Physics, Sobolev Institute of Mathematics,\\
Novosibirsk, Russia
\\ 
$^{d}$ National Research Novosibirsk State University,\\
 Novosibirsk, Russia;\\ pletnev@math.nsc.ru
}
\vspace{.2cm}
\end{center}

\begin{abstract}
We consider the off-shell formulation of the 5D, $\cN=1$ super
Yang-Mills and super Chern-Simons theories in harmonic superspace.
Using such a formulation we develop a manifestly supersymmetric and
gauge invariant approach to constructing the one-loop effective
action both in super Yang-Mills and super Chern-Simons models. On the base of
this approach we compute the leading low-energy quantum contribution
to the effective action on the Abelian vector multiplet background.
This contribution corresponds to the `$F^4$' invariant which is given in
5D superfield form.

\vspace{1cm}
\hspace{6cm}{\it Dedicated to
the memory of Boris Zupnik}

\end{abstract}

\section{Introduction}

The study of the quantum structure of the supersymmetric
five-dimensional field theories attracts recently considerable 
attention, mainly due to attempts to find the effective world-volume
action for multiple M5-branes \cite{bag}.  It was conjectured in \cite{witten} that  the  six-dimensional
(2,0) superconformal field theories on a stack of
M5-branes are equivalent to five-dimensional
super Yang-Mills (SYM) theories. To establish this correspondence, the Kaluza-Klein
reduction of the general 6D (1,0) pseudo-action with a non-Abelian
gauge group $G$ was performed in the series of  papers
\cite{3} and the  5D effective action for the Kaluza-Klein
zero-modes was derived. This duality serves as an important constraint on the
models for multiple M5-branes.  Of course, not all consistent
five-dimensional theories arise in such a circle compactification.

On the other hand, the 5D theory has a global $U(1)$  symmetry and
the current $j=\star F \wedge F$ is always conserved. The
corresponding conserved charge is the instanton number. Such a
conserved current can be coupled to vector superfield what allows us
to identify the scalar component $\phi$ of this vector superfield as
the gauge coupling $<\phi>\sim \f{1}{g^2_{SYM}}$ \cite{seib96}.
Using this observation the authors of papers \cite{M5D5} proposed
that the maximally extended 5D supersymmetric gauge theory describes
the 6D, $(2,0)$ superconformal field compactified on circle without
introducing the Kaluza-Klein reduction. This proposition was based
on the observation that the Kaluza-Klein momentum along the circle
can be identified with instanton charge in the 5D theory. The latter
is a topological charge carried by soliton configurations, which are
analogous to monopole and dyon  configurations  in 4D.  Such an
attractive hypothesis means, in fact, that adding the 1/2-BPS
particle soliton states with instanton number $k$ and a mass formula
$M\sim \f{4\pi k}{g^2_{SYM}}\sim \f{k}{R_5}$ to 5D SYM theory gives
us the full nonperturbative particle spectrum and determines the
nonperturbative completion of the theory under consideration. This
might be an argument in favor of UV finiteness of 5D SYM
perturbative theory and thus be an argument for the consistent
quantum theory. In the strong coupling limit the 5D SYM should
define the fully decompactified 6D, $(2,0)$ theory, which, in its
turn, is expected to describe the low-energy dynamics of multiple
M5-branes. 

An additional important motivation to study  5D
supersymmetric gauge theories comes from the existence of the
corresponding super Chern-Simons (SCS) theory. This theory is 
interesting since it has a conformal fixed point in five dimensions
and can admit a holographic duality \cite{seib96}. There are several reasons why the 5D supersymmetric  Chern-Simons
theory can be interesting in quantum domain. First of all, the
Chern-Simons terms can be generated by integrating out the massive
hypermultiplets in the SYM theory when the hypermultiplets transform
in complex representations of the gauge group \cite{witten}. If we consider the
masses of the hypermultiplets as the UV cut-off, then this leads to
the generation of the Chern-Simons term in the one-loop correction
to the classical theory. Hence inclusion of the SCS term into the
action can be useful in some cases if we want to have a complete
description of the theory. SCS theory can also be important in the
relationship between 5D SYM and 6D, (2,0) theories. In particular,
one can argue \cite{scs5D} that the 5D Chern-Simons term can be
generated by the anomaly terms in the six-dimensional theory. By
focusing on a certain class of anomaly-free six-dimensional theories
the authors of \cite{scs5D} formulated the explicit constraints on
the spectrum and supersymmetry content of the six-dimensional theory
in terms of the five-dimensional Chern-Simons couplings. Therefore
it would be interesting to compute the perturbative quantum
corrections in such 5D theories. In particular, it was demonstrated
that massive fermions running in the loop generate constant
corrections to the 5D Chern-Simons terms of the form $k_{ABC}A^A
\wedge F^B \wedge F^C+\kappa_A A^A \wedge \tr(R\wedge R)$, where
$A^A$  denotes collectively the graviphoton and the vectors from the
vector multiplet,  $F^A$ are the corresponding field strengths, and
$R$ is the curvature two-form.

Though
5D and 3D Chern-Simons theories share some interesting properties,
such as quantization of the level $k$, nevertheless there are also
some differences. The most important difference is the presence of
local degrees of freedom in the higher dimensional case. This
peculiar fact makes it attractive and interesting to perform a more
detailed analysis of 5D SCS theories.

It is known that   non-trivial observables exist in supersymmetric
gauge theories which are not very sensitive to details of the UV
completion. Quantum effects which non-trivially contribute to such
BPS observables are often highly constrained. With using the procedure of localization of the path integral were studied of the various observables in 5D supersymmetric theories \cite{loc}. It was
shown that the partition function for the maximally extended SYM on
$S^5$ captures the physical aspects of the 6D, $(2, 0)$ theory in a
surprisingly accurate and detailed manner. In particular, the $N^3$
behavior of this partition function in 5D supersymmetric gauge
theory is in agreement with the important results obtained for 6D,
$(2,0)$ theory from the supergravity duals and conformal anomaly
\cite{N3}.

The study of  integral invariants in  half-maximally and maximally
extended supersymmetric theories such as  supergravity and SYM
attracts an attention because they can be viewed, on the one hand,
as possible higher order corrections to the string or brane
effective actions and, on the other hand, as  quantum field
counterterms.  It is  well known on the base of the power counting
arguments that the 5D SYM is perturbatively non-renormalizable.
Therefore we should expect an infinite number of divergent
structures at any loop what leads to infinite number of
counterterms. However, the superspace arguments and the requirements
of on-shell supersymmetry  rule out the first divergences in
D-dimensional SYM theory. Actually the divergences can appear at L
loops where D=4+6/L\footnote{For a discussion of this issue see
\cite{Iinv}, \cite{IIinv} and references therein.}. Construction of
the various supersymmetric, gauge invariant functionals in quantum
field theory is conveniently formulated in the framework of the
effective action. The low-energy effective action can be represented
as a series in supersymmetric and gauge invariants with some
coefficients. In general, the supersymmetry together with conformal
symmetry imposes rigid constraints on  these coefficients. In
some cases, they can be determined exactly \cite{dine97}. For
example, the leading term in the low-energy effective action on the
Abelian vector field background is $F^4$ which is generated only at
one loop and is not renormalized at higher loops. A possible new
non-renormalization theorems for Abelian $F^n$ was conjectured in
\cite{non-ren}. Recently the authors of \cite{D6 non-ren}
systematically analyzed the effective action on the moduli space of
(2,0) superconformal field theories in six dimensions, as well as
their toroidal compactification to maximally SYM theories in five
and four dimensions. They  presented an approach to non-renormalization
theorems that constrain this effective action. The first several
orders in its derivative expansion are determined by a one-loop
calculation in five-dimensional SYM theory. In general, the
functional form of the effective action at the first several orders
in the derivative expansion can be obtained by integrating out the
massive degrees of freedom in the path integral. However, it is
difficult enough to perform exactly such an analysis for
supersymmetric models in the component formulation.

Construction of the background field method in
extended supersymmetric gauge theories faces a fundamental problem.
The most natural and proper description of such theories should be
formulated in terms of a suitable superspace and unconstrained
superfields on it. Some time ago a systematic background field method to study the
effective actions in 4D, $\cN = 2$ supersymmetric field theories was
developed in a  series of papers \cite{backgr}, \cite{KmcA}. This method is based
on formulation of $\cN = 2$ theories in harmonic superspace
\cite{gikos}, \cite{gios} and guarantees the manifest $\cN = 2$ supersymmetry and
gauge invariance at all stages of calculations. The method under
consideration gives the possibility to calculate in a straightforward manner not
only the holomorphic and non-holomorphic contributions to the low-energy
effective action but also to study the full structure of the effective
action. Evaluation of the effective action within the background
field method is often accompanied by using the proper time or heat
kernel techniques. These techniques allow us to sum up efficiently
an infinite set of Feynman diagrams with increasing number of
insertions of the background fields and to develop a background
field derivative expansion of the effective action in a manifestly
gauge covariant way. 

The 5D SCS theories are  superconformally invariant and, hence,
their effective actions must be independent of any scale. Unlike the
4D, $\cN=2,4$ supersymmetric theories where  holomorphy allows one to
get the chiral contributions to effective action \cite{seib}, in the
5D case the contributions to the effective action can be written
only either in full or  in analytic superspaces. Then, taking
into account the mass dimensions of the harmonic potential $V^{++}$ and
superfield strength $\cW$ as well as the dimensions of the
superspace measures $d^5x d^4\t^+$ and $d^5x d^8\t$, one obtains
that the most general low-energy effective U(1)-gauge invariant action in the
analytic superspace is the SCS action \cite{z99}, \cite{khyper}. The
next-to-leading effective Abelian 5D action can be written only at
full superspace in terms of the manifestly gauge invariant
functional $\G=\int d^5x d^8\t \ \cW\ {\cal H}(\f{\cW}{\Lambda}),$
where $\Lambda$ is some scale and ${\cal H}(\f{\cW}{\Lambda})$ is
the dimensionless function of its argument. The requirement of scale
invariance means an equation $\Lambda\f{d}{d\Lambda}\int d^5z d^8\t
\ \cW{\cal H}(\f{\cW}{\Lambda})=0$, where the only solution is ${\cal H}=c\ln
\f{\cW}{\Lambda}~.$ Any perturbative or non-perturbative quantum
corrections should be included into a single constant c. The
component Lagrangian in the bosonic sector corresponding to the above
effective action is $\f{1}{{\phi}^3}(F^4+(\p\phi)^4+\l)$, where $F$ is the Abelian
strength of the component vector field from 5D, $\cN=1$ vector multiplet
and ${\phi}$ is the corresponding scalar component.

In this paper we derive the leading contribution to the low-energy
effective action in the 5D SCS theory using the harmonic superspace
description of the theory and proper-time techniques. The result
precisely corresponds to the above analysis and has the form $\int
d^5x d^8\t \ \cW\ \ln\f{\cW}{\Lambda}$. Besides the effective action
in the 5D SCS theory, we calculate the leading one-loop low-energy
contribution in the 5D SYM theory. Although this theory is not
superconformal and  is characterized by the dimensional coupling
constant, its leading contribution to the effective action has the
same functional form as in the 5D SCS theory and does not depend on
the scale and coupling constant. Also, we consider the effective
action in the 5D hypermultiplet theory coupled to a background 5D
vector multiplet. The leading low-energy contribution to effective
action in this theory was calculated in paper \cite{khyper}, where
it was shown that this contribution is the 5D SCS action. In the
given paper we calculated the first next-to-leading term in the
low-energy effective action for the theory under consideration and
found that this term again has the same functional form as the
leading term in 5D SCS theory.

The paper is organized as follows. Section 2 is devoted to a brief
review of harmonic superspace formulation of the 5D, $\cN=1$
supersymmetric field models such as the SYM theory, the
hypermultiplet theory and the SCS theory. In section 3 we consider
the Abelian 5D SCS theory
and develop the manifestly supersymmetric and
gauge invariant procedure for calculating the effective action. This
procedure is based on the background field method and proper-time
technique. We find the  exact expression for one-loop
effective action in terms of functional determinants of differential
operators in analytic subspace of harmonic superspace and calculate
the leading low-energy contribution to this effective action. In section 4 we develop the analogous procedure for
5D SYM theory and calculate the leading low-energy contribution to one-loop effective
action. Section 5 is devoted to the study of the first next-to-leading contribution to effective action in the 5D hypermultiplet
theory coupled to a 5D vector multiplet background. The last section is
devoted to the summary of the results.

\section{Review of the 5D, $\cN=1$ harmonic superspace approach}

Various supersymmetric theories with eight supercharges admit the
off-shell superfield formulations in terms  of formalism of the harmonic superspace.
The harmonic superspace approach for the 4D, $\cN$ = 2 theories was
originally developed in \cite{gikos}. The formulation for the 5D,
$\cN$ = 1 models has been given in  \cite{z99}, \cite{SKD5},
\cite{Kall}. The harmonic superspace approach for the 6D, $ (1,0)$
SYM theories was considered in \cite{howe83}, \cite{ISZ},
\cite{BP15} and for the 6D, $ (1,1)$ SYM in \cite{BIS15}. The
construction the super-de Rham complex in five-dimensional, $\cN$ =
1 superspace and its relationship to the complex of six-dimensional,
$\cN$ = (1, 0) superspace via dimensional reduction was considered
in \cite{gat}. All these harmonic superspace formulations in
space-time dimensions 4, 5 and 6 look almost analogous modulo some
details.  In a series of papers \cite{K0}, \cite{K}, \cite{KN} an
extensive program of constructing the manifestly supersymmetric
formulation for the 5D, $\cN=1$ supergravity-matter models was
realized and the universal procedures to construct manifestly 5D
supersymmetric action functionals for a different supermultiples
were developed. These superfield results are in agreement with the
earlier component considerations of the same models \cite{5Dcomp},
\cite{kugo}.

The study of the structure of the low-energy effective action in the 5D
superconformal theories looks useful from the point of view of the
classification of  theories consistent at the quantum
level. In ref \cite{khyper} a manifestly 5D, $\cN=1$ supersymmetric,
gauge covariant formalism for computing of the one-loop effective
action for a hypermultiplet coupled to a background vector multiplet
was developed. It was demonstrated, as a simple application, that
the SCS action is generated at the quantum level on the Coulomb
branch. The above paper was the only one where explicit one-loop
harmonic superspace calculations were done. We believe that until now
the possibilities of the covariant 5D harmonic superspace methods
were non-sufficiently explored to study the effective action in
5D supersymmetric gauge theories. The aim of this paper is to
develop the general manifestly supersymmetric and gauge invariant
methods for 5D quantum supersymmetric gauge theories and apply these
methods to calculate  the low-energy effective action in  5D
SCS and SYM theories.

In this section we briefly review the superspace description of a
vector multiplet in 5D supersymmetric gauge theories. Our aim
here is to fix our basic notation and conventions (for details
see \cite{SKD5}).

The 5D gamma-matrices $\G_{\hat{a}}$ are defined as follows
$\{\G_{\hat{a}},\G_{\hat{b}}\}=-2\eta_{\hat{a}\hat{b}}{\bf 1},$ with
$\eta_{\hat{a}\hat{b}}=\{-1,1,\l,1\}.$ The matrices $\{{\bf 1},
\G_{\hat{a}}, \S_{\hat{a}\hat{b}}\}$ form a basis in the space of
$4\times 4$ matrices. The charge conjugation matrix
$C=(\ve^{\hat\a\hat\b})$ and its inverse
$C^{-1}=(\ve_{\hat\a\hat\b})$ are used to raise and lower the spinor
indices. The matrices $\ve_{\hat\a\hat\b}$ and
$(\G_{\hat{a}})_{\hat\a\hat\b}$ are antisymmetric, while the
matrices $(\S_{\hat{a}\hat{b}})_{\hat\a\hat\b}$ are symmetric. The anticommuting variables $\t^{\hat\a}_i$ are assumed to obey the
pseudo-Majorana reality condition
$(\t^{\hat\a}_i)^\star=\t^i_{\hat\a}=\ve_{\hat\a\hat\b}\ve^{ij}\t^{\hat\b}_j$.
5D, ${\cal N}=1$ superspace is parameterized by the coordinates
$z^{M}=(x^{\hat\a\hat\b},\t^{\hat\a}_i)$ where $i=1,2$ and
$x^{\hat\a\hat\b}=(\G_{\hat{a}})^{\hat\a\hat\b}x^{\hat{a}}$.

The basic spinor covariant derivatives of the 5D, $\cN=1$ superspace
are
$D^i_{\hat\a}=\f{\p}{\p\t^{\hat\a}_i}-i\p_{\hat\a\hat\b}\t^{\hat\b
i}.$ They obey the anti-commutation relations
\be\label{DD}\{D_{\hat\a}^i,D_{\hat\b}^j\}=-2i\ve^{ij}\p_{\hat\a\hat\b}.\ee

The $\cN=1$ harmonic superspace $R^{5|8}\times S^2$ extends the
conventional 5D, $\cN=1$ Minkowski superspace $R^{5|8}$ with the
coordinates $z^M=(x^{\hat{m}},\t^{\hat\a}_i),$ by the two-sphere
$SU(2)/U(1)$ parameterized by harmonics, i.e., group elements
$$u^\pm_i \in SU(2),\q \overline{{u^{+i}}}=u^-_i, \q
u^{+i}u^-_i=1~.$$ The main conceptual advantage of harmonic
superspace is that the $\cN=1$  vector multiplet as well as
hypermultiplets can be described by unconstrained superfields on
the analytic subspace of $R^{5|8}\times S^2$ parameterized by the
coordinates $\zeta^M=(x^{\hat{m}}_A, \t^{+\hat\a}, u^\pm_i),$ where
\be\label{anbas}
x^{\hat{a}}_A=x^{\hat{a}}+i\t^{+\hat\a}\G^{\hat{a}}_{\hat\a\hat\b}\t^{-\hat\b},
\q \t^\pm=u^\pm_i\t^i.\ee The important property of the coordinates
$\zeta^M$ is that they form a subspace closed under $\cN=1$
supersymmetry transformations. In the coordinates $\zeta^M$ the
spinor covariant derivatives $D^+_{\hat\a}=u^+_i D^i_{\hat\a}$ have
a short form (see Eqs (\ref{ander})) and therefore the superfield
$\Phi(x^{\hat{a}},\t^{\pm\hat\a}, u^\pm_i)$ satisfying the
constraints $D^+_{\hat\a}\Phi=0$ is an analytic superfield
$\Phi(\zeta, u).$ It leads to reducing the number of the anticommuting
coordinates and, hence, to reducing the number of independent
components in comparison with general superfields. However, all
component fields depend now on extra bosonic coordinates $u^\pm_i$.

In harmonic superspace a full set of gauge covariant derivatives
includes the harmonic derivatives which form a basis in the space of
left-invariant vector fields of $SU(2)$: \be D^{++}=u^{+i}\f{\p}{\p
u^{-i}}, \q D^{--}=u^{-i}\f{\p}{\p u^{+i}},\q D^{0}=u^{+i}\f{\p}{\p
u^{+i}}-u^{-i}\f{\p}{\p u^{-i}},\ee \be\label{su2}[D^0,
D^{\pm\pm}]=\pm 2D^{\pm\pm},\q [D^{++},D^{--}]=D^0.\ee
The generator of the $SU(2)$ algebra $D^0$ is an operator of harmonic charge, $D^0\Phi^{(q)}=q\Phi^{(q)}$. In the analytic
basis, parameterized by the coordinates $\zeta^M$ (\ref{anbas})  the
spinor covariant derivatives: $D^\pm_{\hat\a}=u^\pm_i D^i_{\hat\a}$
and the harmonic derivatives take the form \be\label{ander}
D^+_{\hat\a}=\f{\p}{\p\t^{-\hat\a}},\q
D^-_{\hat\a}=-\f{\p}{\p\t^{+\hat\a}}-2i\p_{\hat\a\hat\b}\t^{-\hat\b},\ee
$$D^{++}=u^{+i}\f{\p}{\p u^{-i}}+i\t^{+\hat\a}\p_{\hat\a\hat\b}\t^{+\hat\b}+\t^{+\hat\a}\f{\p}{\p\t^{-\hat\a}},
\q D^{--}=u^{-i}\f{\p}{\p
u^{+i}}+i\t^{-\hat\a}\p_{\hat\a\hat\b}\t^{-\hat\b}+\t^{-\hat\a}\f{\p}{\p\t^{+\hat\a}},$$
$$D^0=u^{+i}\f{\p}{\p u^{+i}}-u^{-i}\f{\p}{\p u^{-i}}+\t^+\f{\p}{\p\t^+}-\t^-\f{\p}{\p\t^-}.$$
In this basis Eqs (\ref{DD}), (\ref{su2}) leads to
\be\label{covalg2}\{D^+_{\hat\a},D^+_{\hat\b}\}=0, \q [D^{\pm\pm}, D^\pm_{\hat\a}]=0,\q
[D^{\pm\pm},D^\mp_{\hat\a}]=D^{\pm}_{\hat\a},\ee
$$\{D^+_{\hat\a},D^-_{\hat\b}\}=2i\p_{\hat\a\hat\b}\,.$$
These relations are necessary integrability conditions for the
existence of the analytic superfields. Since the field in the
harmonic superspace depend on the additional bosonic variable $u^i$,
we must define rules of integration over harmonics (that is, over the group
manifold $SU(2)/U(1)$). The basic harmonic integrals have the form
\cite{gios} \be\int du =1, \q \int du u^+_{(i_1}\l
u^+_{i_n}u^-_{j_1}\l u^-_{j_m)}=0, \q n+m
>0\,.\ee It means that the harmonic integrals are nonzero only if the integrand
has the zero $U(1)$ charge.

\subsection{5D SYM theory in harmonic superspace}
To describe a Yang-Mills supermultiplet in 5D conventional
superspace we introduce the gauge covariant derivatives
$\cD_A=D_A+i\cA_A$ where $D_A=(\p_{\hat{a}}, D^i_{\hat\a})$ are the
flat covariant derivatives  and $\cA_A$ is the gauge connection
taking values in the Lie algebra of the gauge group. The operator
$\cD_A$ satisfies the gauge transformation law $\cD_A\rightarrow
e^{i\tau(z)}\cD_A e^{-i\tau(z)},\q \tau^\dagger =\tau$ with a
superfield gauge parameter $\tau(z)$. The gauge covariant
derivatives are required to obey some constraints \cite{HL}
\be\label{alg} \{\cD_{\hat\a}^i, \cD_{\hat\b}^j\}=-2i
\ve^{ij}(\cD_{\hat\a\hat\b}+\ve_{\hat\a\hat\b}i\cW),\ee
$$[\cD^i_{\hat\a}, \cD_{\hat{a}}]=-i(\G_{\hat{a}})_{\hat\a}^{\ \hat\b}\cD^i_{\hat\b} \cW,
\q
[\cD_{\hat{a}},\cD_{\hat{b}}]=-\f14(\S_{\hat{a}\hat{b}})^{\hat\a\hat\b}\cD^i_{\hat\a}\cD_{\hat\b
i}\cW=i{\cF}_{\hat{a}\hat{b}},$$ with the matrices $\G_{\hat{a}}$
and $\S_{\hat{a}\hat{b}}$ defined above\footnote{The properties of
these matrices are discussed in \cite{SKD5}, \cite{K}.}. Here the
field strength $\cW$ is Hermitian, $\cW^\dagger=\cW$ and obeys the
Bianchi identity \be\label{Bianch}
\cD_{\hat\a}^{(i}\cD_{\hat\b}^{j)}\cW=\f14\ve_{\hat\a\hat\b}\cD^{\hat\g(i}\cD^{j)}_{\hat\g}\cW.\ee

In harmonic superspace the full set of gauge covariant derivatives includes the harmonic derivatives.
In this basis Eqs. (\ref{alg}), (\ref{ander}) leads to
\be\label{covalg}\{\cD^+_{\hat\a},\cD^+_{\hat\b}\}=0, \q [D^{\pm\pm}, \cD^\pm_{\hat\a}]=0,\q
[D^{\pm\pm},\cD^\mp_{\hat\a}]=\cD^{\pm}_{\hat\a},
\ee
$$\{\cD^+_{\hat\a},\cD^-_{\hat\b}\}=2i\cD_{\hat\a\hat\b}-2\ve_{\hat\a\hat\b}\cW, \q
[\cD^\pm_{\hat\g},\cD_{\hat\a\hat\b}]=i\{\ve_{\hat\a\hat\b}\cD^\pm_{\hat\g}\cW+2\ve_{\hat\g\hat\a}\cD^\pm_{\hat\b}\cW-
2\ve_{\hat\g\hat\b}\cD^\pm_{\hat\a}\cW\}~,$$
$$i{\cF}_{\hat{a}\hat{b}}=\f12 \cD^-_{\hat\a}(\S_{\hat{a}\hat{b}})^{\hat\a\hat\b}\cD^+_{\hat\b}\cW~.
$$
 The integrability
condition $\{\cD^+_{\hat\a},\cD^+_{\hat\b}\}=0$ is solved by
$\cD^+_{\hat\a}=e^{-iv}D^+_{\hat\a} e^{iv}$ with some Lie-algebra
valued harmonic superfield $v(z,u)$ which is called the bridge. The
bridge possesses a richer gauge freedom than the original
$\tau$-group \be e^{iv'(z,u)}=e^{i\lambda(z,u)}e^{iv}e^{-i\tau(z)},
\ D^+_{\hat\a}\lambda=0.\ee The $\lambda$- and $\tau$-
transformations generate, respectively, the so-called $\lambda$- and
$\tau$-groups. In the $\lambda$-frame the spinor covariant
derivatives $\cD^+_{\hat\a}$ coincide with the flat ones, while the
harmonic covariant derivatives acquire connections
$\cD^{\pm\pm}=D^{\pm\pm} +i V^{\pm\pm}$. The real connection
$V^{++}$ is  an analytic superfield, $D^+_{\hat\a} V^{++}=0$ and its
transformation law is \be\label{gaugetr}V^{++\
'}=e^{i\lambda}V^{++}e^{-i\lambda}-i
e^{i\lambda}D^{++}e^{-i\lambda}.\ee The gauge freedom $-\d
V^{++}=\cD^{++}\lambda$ can be used to impose the Wess-Zumino gauge
in the form \be\label{Vcomp}V^{++}=(\t^+)^2
i\phi+i\t^{+\hat\a}A_{\hat\a\hat\b}\t^{+\hat\b}+4(\t^+)^2\t^{+\hat\a}\lambda^-_{\hat\a}-\f32(\t^+)^2(\t^+)^2Y^{--}.\ee
In this gauge, the superfield $V^{++}$ contains the real scalar
field $\phi$, the Maxwell field $A_{\hat{m}}$, the isodoublet of
spinors $\lambda_{\hat\a}^i$ and the auxiliary isotriplet
$Y^{(ij)}$. The analytic superfield $V^{++}$ turns out to be the
single unconstrained potential of the pure 5D, $\cN = 1$ SYM theory
and all other quantities, associated with this theory, are expressed
in its terms. In particular, the other harmonic connection $V^{--}$
turns out to be uniquely determined in terms of $V^{++}$ using the
zero-curvature condition $[\cD^{++},\cD^{--}]=D^0.$ The result looks
like \cite{Z}
\be\label{V^--}V^{--}(z,u)=\sum_{n=1}^\infty(-i)^{n+1}\int du_1\l
du_n\f{V^{++}(\zeta,u_1)\l V^{++}(\zeta
,u_n)}{(u^+u^+_1)(u^+_1u^+_2)\l(u^+_nu^+)},\ee where
$(u^+_1u^+_2)=u^{+i}_1u^{+}_{2  i}$. The details of the harmonic
analysis  on $S^2=SU(2)/U(1)$) were designed in the pioneering works
\cite{gikos}, \cite{graph}, \cite{Z}, \cite{gios}. In the
$\lambda$-basis the connections $\cA^-_{\hat\a}$, $\cA_{\hat{m}}$
and the field strength can be expressed in terms of $V^{--}$ using
the relations (\ref{covalg}): \be \cA^-_{\hat\a}=-D^+_{\hat\a}
V^{--}, \q \cW_\lambda=\f{i}{8}D^{+\hat\a}D^+_{\hat\a} V^{--},\q
{\cF}_{\hat\a\hat\b}=-\f{i}{2}\cD^-_{(\hat\a}\cD^+_{\hat\b)}\cW.\ee
The superfield strength $\cW_\lambda$ satisfies the following
constraints
$\cD^{++}\cW_\lambda=D^{++}\cW_\lambda+i[V^{++},\cW_\lambda]=0.$ In
the Abelian case, the superfield $\cW_\lambda=\f{i}{8}\int du(D^-)^2
V^{++}$ does not depend on harmonics and moreover, in this case
there is no distinction between $\cW_\lambda$ and $\cW$. The Bianchi
identity (\ref{Bianch}) takes the form
\be\cD^+_{\hat\a}\cD^+_{\hat\b}
\cW=\f14\ve_{\hat\a\hat\b}(\cD^+)^2\cW,\ee where we have used the
notation $(\cD^+)^2=\cD^{+\hat\a}\cD^+_{\hat\a}.$ Using these
identities, the authors of \cite{SKD5} built an important
covariantly analytic superfield ${\cal G}^{++}$:
\be\label{calG}-i{\cal
G}^{++}=\cD^{+\hat\a}\cW\cD^+_{\hat\a}\cW+\f14\{\cW,(\cD^+)^2\cW\},\q
\cD^+_{\hat\a}{\cal G}^{++}=0, \q \cD^{++}{\cal G}^{++}=0~.\ee
 This superfield can be transformed to the form
\be{\cal G}^{++}= (D^+)^4(V^{--}\cW),\ee where $
(\cD^\pm)^4=-\f{1}{32}(\cD^\pm)^2(\cD^\pm)^2$ and
$\f{1}{2}(D^+)^4(D^{--})^2\Phi(\zeta)=
-\f14\cD^{\hat\a\hat\b}\cD_{\hat\a\hat\b}\Phi(\zeta).$

Unlike the 4D, $\cN=2$ case, the chiral superspaces are not
Lorentz-covariant in the case of 5D, so that to construct the
superfield actions we can use only the full superspace or else the
analytic superspace. In the full harmonic superspace the 5D SYM
action has the universal form \cite{Z} \be\label{SYM}
S_{SYM}(V^{++})=\f{1}{g^2_{SYM}}\sum_{n=1}^\infty\f{(-1)^n}{n}\tr\int
d^{13}z du_1\l du_n\f{V^{++}(z,u_1)\l V^{++}(z,u_n)}{(u^+_1 u^+_2)\l
(u^+_n u^+_1)}~, \ee where $g^2_{SYM}$ is the coupling constant of
dimension $[g^2_{SYM}]=1$. The SYM action in terms of the component
fields defined by (\ref{Vcomp}) is \be S_{SYM}=\f{1}{g^2_{SYM}}\int
d^5x\{-\f14
F^{ab}F_{ab}-\f12\cD^a\phi\cD_a\phi+\f14Y^{ij}Y_{ij}+\f{i}{2}\lambda^i\not\!\!\cD\lambda_i-\f12\lambda^i[\phi,\lambda_i\}.\ee
The SYM equations of motions $(D^+)^2\cW=0,\q \Box \cW=0$ have a
vacuum Abelian solution $V^{\pm\pm}=i(\t^\pm)^2 Z$ where $Z$ is the
linear combination of the Cartan generators of the gauge group
\cite{z99}. This vacuum solution spontaneously breaks the gauge
symmetry, but it conserves the 5D supersymmetry with the central
charge  and produces BPS masses of the Z-charged fields. In addition,
let us note  that on-shell, the degrees of freedom $(1+3)_B+4_F$
in  $\cW$ match the degrees of freedom in a 5D tensor
multiplet described by the  superfield $\Phi $ subject to the constraint
$D^{(i}_{\hat\a} D^{j)}_{\hat\b}\Phi=0.$ The $\t$-expansion of
$\Phi$ is \be\Phi=\varphi +\t^{\hat\a}_i\chi^i_{\hat\a} +\t^{\hat\a\
i}\t^{\hat\b}_i H_{\hat\a\hat\b}+\l,\ee where
$H_{\hat\a\hat\b}=\f12(\S^{\hat{a}\hat{b}})_{\hat\a\hat\b}H_{\hat{a}\hat{b}}$
is dual to the 3-form field strength $F_{\hat{a}\hat{b}\hat{c}}$ of
the 2-form gauge field $B_{\hat{a}\hat{b}}$. In this situation  the
vector massless representation is equivalent under duality to a
tensor representation.

The other universal procedure to construct 5D manifestly
supersymmetric actions is based on ideas developed in \cite{sohn}.
Let us consider two vector multiplets, namely a $U(1)$ vector multiplet
$V_\Delta^{++}$ and a Yang-Mills vector multiplet $V^{++}_{SYM}$.
They can be coupled in a gauge-invariant way, using the interaction
\be\label{actYM}S_{SYM}=\int d\zeta^{(-4)}V^{++}_\Delta\tr {\cal
G}^{++}_{SYM},\ee
 where ${\cal G}^{++}_{SYM}$ corresponds to a
non-Abelian multiplet and is defined in eg. (\ref{calG}). If we
assume that the physical scalar field in $V^{++}_\Delta$ possesses a
non-vanishing expectation value
$<V^{++}_\Delta(\zeta,u)>=i(\t^+)^2\f{1}{g^2_{YM}}$, then one gets
\be\label{actYM-2} S_{SYM}=\f{1}{g^2_{YM}}\int
d\zeta^{(-4)}i(\t^+)^2\tr{\cal G}^{++}_{SYM}.\ee The integration in
(\ref{actYM}), (\ref{actYM-2}) is carried out over the analytic
subspace of the harmonic superspace \be\int d\zeta^{(-4)}\equiv\int
d^5x du (\cD^-)^4. \ee

\subsection{The super Chern-Simons model}
Let us consider the 5D harmonic superspace formulation of a
supersymmetric Chern-Simons (SCS) model. The off-shell non-Abelian
SCS action  in five dimensions was first constructed in \cite{kugo}.
SCS model possesses remarkable properties, which makes this theory
very interesting for various applications. First of all, in five
dimensions the pure $U(N)$ and $SU(N)$ Chern-Simons theories have
superconformal fixed points \cite{seib96}. Another reason to study
the SCS models is that they can be generated by integrating out the
massive hypermultiplets in the SYM theory \cite{witten},
\cite{scs5D} when the hypermultiplets transform in complex
representations of the gauge group. For example, if we consider the
masses of the hypermultiplets as the UV cut-off, we obtain the
Chern-Simons term in the one-loop correction to the classical
theory. Hence inclusion of this term can be important in some cases
to get a complete description of the theory. In a manifestly
supersymmetric setting, where the entire vector supermultiplet is
taken into account, the corresponding one-loop calculation was given
in \cite{khyper}, both in the Coulomb and non-Abelian phases. Using
the covariant harmonic supergraphs and the heat kernel techniques in
harmonic superspace \cite{backgr}, \cite{KmcA}, it was shown that
the hypermultiplet effective action contains the SCS term. Finally
Chern-Simons theory can be important in the relation between the 5D
SYM and 6D $(2,0)$ theories. In particular, one can argue
\cite{scs5D} that the 5D Chern-Simons term can be generated by the
anomaly terms in the six-dimensional theory.

In general, in space-time with dimension $(2n-1)$, the
action of the Chern-Simons theory can be constructed using the Chern-Simons
form $\Sigma_{2n-1}$, defined as
$d\Sigma_{(2n-1)}=\tr[F^n]$
where $F = d\cA+i\cA\wedge \cA$ is the gauge field strength two-form and its powers $F^n$
are defined by the  wedge product. In three space-time dimensions this form gives
rise to the famous 3D Chern-Simons action. In the 5D Chern-Simons theory the action is given by
$$S_{CS} = \f{k}{12}\int d^5x  \ve^{\hat{a}\hat{b}\hat{c}\hat{d}\hat{e}}\tr\{\cA_{\hat{a}}
F_{\hat{b}\hat{c}}F_{\hat{d}\hat{e}}-i\cA_{\hat{a}} F_{\hat{b}\hat{c}}F_{\hat{d}\hat{e}}-
\f25\cA_{\hat{a}}\cA_{\hat{b}}\cA_{\hat{c}}\cA_{\hat{d}}\cA_{\hat{e}}\},
$$
where $k$ is the Chern-Simons level which plays the role of the
coupling constant. Here $\cA_{\hat{a}}$ is the gauge field with the
gauge group $SU(N)$ or $U(N)$. The field $\cA_{\hat{a}}$ transforms
under the gauge transformation $g$ as $\cA_{\hat{a}}\rightarrow
g^{-1}\cA_{\hat{a}} g-ig^{-1}\p_{\hat{a}} g.$ Under this
transformation, $S_{CS}$ gets an additional term $\d S_{CS}$ given,
modulo a total derivative, by $\d S_{CS}=2\pi i k \ Q(g),$ where
$Q(g)$ takes only integer values. Like in the case of
three-dimensional Chern- Simons theory,  gauge invariance of the
partition function leads to $k\in Z$. However, despite the fact that
the action of the theory CS  does not depend on the metric and thus the
theory is topological,  the five (and, generally, any higher odd)
dimensional case admits  local propagating degrees of freedom
\cite{banados}.

Unlike the component construction of \cite{kugo}, a closed-form
expression for the non-Abelian SCS action has never been given in terms
of the superfields. Here there exists only a unique definition of the
variation of the SCS action with respect to an infinitesimal
deformation of the gauge potential $V^{++}$. However, as it was
noted in \cite{KN}, it is unknown how to integrate this variation in
a closed form to obtain the action as an integral over the
superspace. The component formulation of the non-Abelian SCS model can be
constructed in the framework of the superform approach \cite{KN},
where a closed-form expression for the non-Abelian SCS action was
given. The superfield analysis in the Abelian case is more transparent
and the SCS action was derived in the 5D $\cN = 1$ harmonic
\cite{z99}, \cite{SKD5} and projective superspaces \cite{K}.

The approach of constructing the  manifestly supersymmetric actions
developed in \cite{z99}, \cite{SKD5} leads to Abelian SCS action in
the form \be\label{CSaction} S_{SCS}=\f{1}{12g^2} \int d\zeta^{(-4)}
V^{++}{\cal G}^{++}, \q \cD^+_{\hat\a} {\cal G}^{++}=\cD^{++}{\cal
G}^{++}=0.\ee
 The action (\ref{CSaction}) is
invariant under the gauge transformations $-\d
V^{++}=D^{++}\lambda.$ The equations of motions for the model with
such an action are \be\label{EoM}-i{\cal G}^{++}=\cD^{+\hat\a}\cW
\cD^+_{\hat\a}\cW +\f12 \cW (\cD^+)^2\cW=0.\ee

The Abelian SCS theory with the superfield action (\ref{CSaction}) leads to
the following component action: \be\label{CScomp} S_{SCS}=\f{1}{2g^2}
\int d^5x \{\f13\ve^{\hat{a}\hat{b}\hat{c}\hat{d}\hat{e}}A_{\hat{a}}
F_{\hat{b}\hat{c}}F_{\hat{d}\hat{e}}-\f12\phi
F^{\hat{a}\hat{b}}F_{\hat{a}\hat{b}}-\phi\p^{\hat{a}}\phi\p_{\hat{a}}\phi+\f12\phi
Y^{ij}Y_{ij} \ee
$$-\f{i}{2}F_{\hat{a}\hat{b}}(\psi^i\S^{\hat{a}\hat{b}}\psi_i)+i\phi(\psi^k\not\!\p\psi_k)-
\f{i}{2}Y_{ij}(\psi^i\psi^j)\}.$$ The action (\ref{CScomp}) clearly
shows that the five dimensional Abelian SCS theory
 is a non-trivial interacting field model (see e.g. \cite{smilga}).

The theory (\ref{CScomp}) is superconformal at the classical level
and the coupling constant $g^2$ is dimensionless. The latter can
mean a renormalizability of the theory. However, the action
(\ref{CScomp}) does not involve a quadratic part and
perturbative calculations, based on the weakness of the interaction
term with respect to the free part, are impossible. However, we can
use the presence of the vacuum moduli space $<\phi>=m$ in the
Lagrangian (\ref{CScomp}). This allows one to decompose the
Lagrangian into the free and interaction parts and construct the $S$
matrix in a conventional way. However in this case, conformal
symmetry of the original Lagrangian is broken spontaneously and the
mass parameter in the action makes the theory nonrenormalizable.

\subsection{5D $\cN=1$ hypermultiplet in harmonic superspace}
Here we briefly discuss the hypermultiplet formulation in harmonic
superspace.

On mass-shell, the 5D, $\cN=1$ massless hypermultiplet contains  two
complex scalar fields forming an isodoublet of the automorphism
group $SU(2)_A$ of the supersymmetry algebra (\ref{DD}) and an
isosinglet Dirac spinor field. The description  of the off-shell
hypermultiplet in  terms of the analytical supspace of harmonic
superspapce is completely analogous to the description of a 4D,
$\cN=2$ hypermultiplet. Like in the four-dimensional 4D, $\cN = 2$
supersymmetric theory \cite{gikos}, the off-shell hypermultiplet
coupled to the vector supermultiplet is described by a superfield
$q^+(\zeta)$ and its conjugate $\bar{q}^+(\zeta)$ with respect to
the analyticity preserving conjugation \cite{gios}. The classical
action for a massless hypermultiplet coupled to the background 5D,
$\cN = 1$ vector multiplet is \be S_{hyper}=-\int
d\zeta^{(-4)}\bar{q}^+\cD^{++}q^+.\ee

The hypermultiplet that transforms in a real representation of the
gauge group,  on-shell has  scalars in the representations
$(\underline{1}+\underline{3})$ of $SU(2)_A$ and an $SU(2)_A$
doublet of pseudo-Mayorana fermions. It can be described by a real
analytic superfied $\omega(\zeta)$. Such a superfield is called the
$\omega$-hypermultiplet. The action describing an interaction of
this hypermultiplet with a vector supermultiplet is written in the
form
\be S_{\omega}=-\f12\int d\zeta^{(-4)}(\cD^{++}\omega)^2.\ee

\section{The background field formulation for quantum $\cN=1$ super Chern-Simons }
In this section we will construct the background field method for the
superfield theory with action (\ref{CSaction}).

The harmonic superfield background field method (see construction of
this method for 4D, $\cN=2$ SYM theory in \cite{backgr}) is based on the
so-called background-quantum splitting of the initial gauge field
into two parts: the background field $V^{++}$ and the quantum field
$v^{+}$ \be V^{++}\rightarrow V^{++}+v^{++}.\ee To quantize the
theory, one imposes the gauge fixing conditions only on the quantum
field, introduces the corresponding ghosts and considers the
background field as the functional argument of the effective action.
Then, the original infinitesimal gauge transformations
(\ref{gaugetr}) can be realized in two different ways: as the
background transformations \be\d V^{ ++} = -D^{++}\lambda - i[V^{
++}, \lambda] = -\cD^{++}\lambda, \q \d v^{++} = i[\lambda,
v^{++}]~, \ee and as the quantum transformations \be\d V^{ ++} =
0,\q  \d v^{++} = - \cD^{++}\lambda - i[v^{++}, \lambda] .\ee To
construct an effective action as a gauge-invariant functional of the
background superfield $V^{++}$  we will use another form of writing
the action (\ref{CSaction})\footnote{Here we use the identity
$$-\f{1}{32}(\cD^+)^2(\cD^+)^2[V^{--}\cW]=-\f{1}{32}(\cD^+)^2\{-8i\cW\cW+2\cD^{+\b}V^{--}\cD^+_\b\cW+V^{--}(\cD^+)^2\cW\}$$
$$=i(\cD^{+\a}\cW\cD^+_\a\cW+\f14\{\cW,(\cD^+)^2\cW\})$$}
\be S_{SCS}=\f{1}{12g^2} \tr\int d\zeta^{(-4)} V^{++}{\cal G}^{++}=\f{1}{12g^2} \tr\int d^5x d^8\t du\ \  V^{++}V^{--}\cW_\lambda, 
\ee
and expand the action $S[V^{++}+v^{++}]$ in powers of the quantum field
$v^{++}$: \be\label{quantexp}S=\sum_{n=1}^\infty \tr \int
d^{13}z_1du_1\l dz_n du_n\f{1}{n!}\f{\d^n S}{\d v^{++}(1)\l \d
v^{++}(n)}|_{v^{++}=0}v_\tau^{++}(1)\l v_\tau^{++}(n).\ee Here
$\cW_\lambda$,  and $v^{++}_\tau$ denote the $\lambda$- and
$\tau$-frame forms of $\cW$,  and $v^{++}$ respectively
$$\cW_\lambda=e^{iv}\cW_\tau e^{-iv},\q v^{++}_\tau=e^{-iv}v^{++} e^{iv}.$$
Each term in the action (\ref{quantexp}) is manifestly invariant with respect to the background
gauge transformations.
The first variation of the action is
\be\d S_{SCS}=\f{1}{4g^2}\tr\int d^{13}z du \d V^{++} V^{--}\cW_\lambda=
\f{1}{4g^2}\tr\int d^{13}z du v^{++}_\tau V^{--}_\tau\cW_\tau,
\ee
 where $v^{++}_\tau=e^{-iv}\d V^{++}e^{iv}$. It depends on $V^{++}$ via
the dependence of $v^{++}_\tau$ on the bridge $v$. The term linear in $v^{++}$
 determines the equations of motion.
This term should be dropped when one considers the effective action.

To determine the second variation of the action it is necessary to
express $\d V^{--}$ via $\d V^{++}$. A variation of
$V^{++}=-ie^{iv}(D^{++}e^{-iv})$ can be represented as $e^{-iv}\d
V^{++}e^{iv}=iD^{++}(e^{-iv}\d e^{iv})=\d V^{++}_\tau.$ This
equation is solved in the form \be(e^{-iv}\d e^{iv})=-i\int
du_1\f{(u^+ u^-_1) }{(u^+u^+_1)}\d V^{++}_\tau(u_1).\ee Now \be\d
V_\tau^{--}=iD^{--}(e^{-iv}\d e^{iv})=D^{--}\int du_1\f{(u^+ u^-_1)
}{(u^+u^+_1)}\d V^{++}_\tau(u_1)=\int du_1\f{\d
V^{++}_\tau(u_1)}{(u^+u^+_1)^2}.\ee
Calculating the second variation
of the action yields the result \be\label{delta2a} \d^2
S_{SCS}=\f{1}{2} \int d^5x d^8\t \f{du_1du_2}{(u^+_1u^+_2)^2}\
\f{1}{g^2(\cW)} \ \d V^{++}(u_1)\d V^{++}(u_2)~.\ee Here \be
\f{1}{g^2(\cW)}={\cW}~.\ee These two expressions illustrate the
specific feature of 5D, $\cN=1$ SCS theory as a theory with local
coupling constant (since there is  a non-trivial
background-dependent factor in the vector field kinetic operator)
\cite{backgr}.

In the framework of the background field method, we should fix only
the quantum field gauge transformations. As to the superfield $\cW$
it is invariant under these transformations.  Let us introduce the
gauge fixing function in the form \be F^{(+4)}=D^{++} v^{++},\ee
which changes by the law \be\label{dF}\d F^{(+4)}=
(\cD^{++}(\cD^{++}\lambda+i[v^{++},\lambda])),\ee under the quantum
gauge transformations. Eq. (\ref{dF}) leads to the
Faddeev-Popov determinant and to the
corresponding ghost action
\be S_{FP}=\tr\int d\zeta^{(-4)}du \ {\bf b}\
(\cD^{++})^2{\bf c}.\ee Next we average the $\d(F^{(+4)}-f^{(+4)})$
with the weight
\be\label{weight}1=\Delta[V^{++}]\exp\{\f{1}{2\a}\int d^{13}z du_1
du_2f_\tau^{(+4)}(u_1)\cW\f{(u^-_1u^-_2)}{(u^+_1u^+_2)^3}
f_\tau^{(+4)}(u_2)\}.\ee Here $\a$ is an arbitrary gauge parameter,
$\Delta[V^{++}]$ is determinant of the Nielson-Kallosh ghosts and
$f^{(+4)}_\tau=e^{-iv}f^{(+4)}e^{iv}$ is a tensor of the
$\tau$-group. Note that we need the insertion of the superfield
$\cW$ to balance the dimensions. The functional $\Delta[V^{++}]$
is chosen from the condition
$$\Delta^{-1}[V^{++}]=
\int \cD f^{(+4)}\exp\{\f{1}{2\a}\int
d\zeta^{(-4)}_1d\zeta^{(-4)}_2f_\tau^{(+4)}(1)\hat{A}(1,2)f_\tau^{(+4)}(2)\}=
{\rm Det}^{-1/2}\hat{A}.$$
 The background field dependent operator $\hat{A}$ has the form
$$\hat{A}=\f{(u^-_1u^-_2)}{(u^+_1u^+_2)^3}(D^+_1)^4\cW(1)(D^+_2)^4\d^{13}(z_1-z_2),$$
and act on the space of analytic superfields. Thus \be\Delta[V^{++}]
= {\rm Det}^{1/2}\hat{A}.\ee

To find  ${\rm Det}\hat{A}$ we represent it by a functional
integral over analytic superfields of the form \be{\rm
Det}^{-1}(\hat{A})=\int
\cD\chi^{(+4)}\cD\rho^{(+4)}\exp\big\{-i\tr\int
d\zeta^{(-4)}_1d\zeta^{(-4)}_2\chi^{(+4)}(1)\hat{A}(1|2)\rho^{(+4)}(2)\big\},\ee
and perform the following replacement of functional variables
\be\rho^{(+4)}= (\cD^{++})^2\s,\q\q {\rm Det}(\f{\d\rho^{(+4)}}{\d
\s})={\rm Det} (\cD^{++})^2.\ee Further, repeating the construction
of the effective action from the work \cite{graph} we obtain
$$\int d^{13}z du_1du_2\chi_1^{(+4)}\cW\f{(u^-_1u^-_2)}{(u^+_1u^+_2)^3}(\cD_2^{++})^2\s_2=\int d^{13}z du \chi^{(+4)}_\tau\cW\f12(D^{--})^2\s_\tau$$$$=\int d\zeta^{(-4)}\chi^{(+4)}\sB_\cW \s,$$
where \be\label{sB}\sB_{\cW}=\f12(\cD^+)^4\cW(\cD^{--})^2, \ee is the
deformed version  of the covariant analytic d'Alembertian
\cite{SKD5} \be\label{covbox}\sB=\f12(\cD^+)^4(\cD^{--})^2. \ee

Eventually we get the representation of  $\Delta[V^{++}]$ by the
following functional integral \be\label{Delta}\Delta[V^{++}] ={\rm
Det}^{-1/2}(\cD^{++})^2{\rm Det}_{(4,0)}^{1/2}(\sB_\cW)={\rm
Det}_{(4,0)}^{1/2}(\sB_\cW)\int \cD\varphi e^{iS_{NK}[\varphi,
V^{++}]}.\ee
Here  ${\rm Det}_{(q,4-q)}(H)$ is defined as follows
\be {\rm Det}_{(q,4-q)}\hat{H} = e^{\Tr_{(q,4-q)}\ln\hat{H}}, \ee
 and
 the functional trace of operators acting on the space
of covariantly analytic superfields of $U(1)$ charges $q$ and $4-q$
is
 \be \Tr_{(q,4-q)} \hat{H} = \tr\int d\zeta^{(-4)}{\cal
H}^{(q,4-q)}(\zeta, \zeta), \ee where '$\tr$' is the matrix trace and ${\cal
H}^{(q,4-q)}(\zeta_1, \zeta_2)$  is the kernel of the operator.
 Superfield
$\varphi$ in (\ref{Delta}) is a bosonic real analytic superfield,  the Nielsen-Kallosh ghost,
with the action \be S_{NK}=-\f12\int
d\zeta^{(-4)}\cD^{++}\varphi \ \cD^{++}\varphi.\ee

Upon averaging the effective action $\G_{SCS}[V^{ ++}]$ with the
weight (\ref{weight}), one gets the following path integral
representation of the one-loop effective action $\G_{SCS}^{(1)}[V^{
++}]$ for the Abelian 5D SCS theory \be\label{def}
e^{i\G_{SCS}^{(1)}[V^{++}]}=e^{iS_{SCS}[V^{++}]}\int \cD v^{++}\cD
{\bf b}\cD{\bf c}\cD\varphi e^{i S_{quant}[v^{++},{\bf b}, {\bf
c},\varphi, V^{++}]},\ee where \be\label{defq} S_{quant}= \Delta
S_{SCS}[v^{++},V^{++}] +S_{gh}[v^{++},V^{++}]+S_{FP}[{\bf b}, {\bf
c},v^{++},V^{++}]+S_{NK}[\varphi,V^{++}].\ee Here $S_{gh}[v^{++},
V^{++}]$ is the gauge fixing contribution to the action of quantum
superfields  \be\label{gh}S_{gh}[v^{++}, V^{++}]=\f{1}{2\a}\int
d^{13}zdu_1du_2\cD^{++}_1v^{++}_1\cW\f{(u^-_1u^-_2)}{(u^+_1u^+_2)^3}\cD^{++}_2
v^{++}_2\ee
$$=\f{1}{2\a}\int d^{13}z du_1du_2\cW\f{v^{++}_1v^{++}_2}{(u^+_1u^+_2)^2} +
\f{1}{2\a}\int d^{13}z du_1du_2\cW
v_1^{++}\f12(D^{--})^2\d^{(2,-2)}(u_1,u_2)v^{++}_2.$$ The sum of the
quadratic part $\Delta S_{SCS}$ (\ref{delta2a}) in quantum
superfield $v^{++}$ and $S_{gh}$ (\ref{gh}) for special values
$\a=-1$ has the form \be -\f{1}{2}\int d\zeta^{(-4)}du
v^{++}(\cD^+)^4\cW(\cD^{--})^2v^{++}=-\f{1}{2}\int d\zeta^{(-4)}du
v^{++}\sB_{\cW}v^{++}.\ee Eqs. (\ref{def})-(\ref{defq}) completely
determine the structure of the perturbation expansion to
calculate the one-loop effective action $\G_{SCS}^{(1)}[V^{++}]$
of the pure $\cN = 1$ SCS theory in a manifestly supersymmetric and
gauge invariant form\footnote{We restricted ourselves to the  one loop
approximation. However, the same construction will be valid at any
loop if we replace in (\ref{def}) the quadratic part of the action for the
quantum superfields with the general non-quadratic action.}.
The complete  one-loop effective
action is given by the sum of the contributions coming from the ghost superfields and from the
quantum superfields $v^{++}$. It has the form
\be\label{oneloop}\tilde{\G}^{(1)}_{SCS}={i}\big[\f12\Tr\ln(\cD^{++})^2-\Tr\ln(\cD^{++})^2\big]+\f{i}{2}\big\{\Tr\ln\sB_{\cW
\ (2,2)}- \Tr\ln\sB_{\cW \ (4,0)}\big\}.\ee Here the first term is the
contribution from the Nielsen-Kallosh ghosts, the second term is the
contribution from the Faddeev-Popov ghost and the third term is the
contribution from the vector multiplet.  In the next subsection we will
consider the contribution to this effective action from the SCS
multiplet, The contribution from the ghosts will be considered in
section 5.

\subsection{The proper-time representation of the contributions in (\ref{oneloop}) from
the Chern-Simons vector multiplet}
 The contribution of the Chern-Simons vector multiplet to
(\ref{oneloop}) can be written in the proper-time representation
(see the analogous representation for 4D, $\cN=2$ SYM theories in
\cite{KmcA}): \be\label{propt}\G^{(1)}_{SCS}=-\f{i}{2}\int_0^\infty
\f{ds}{s}\Tr(e^{is\sB_\cW}\Pi_T^{(2,2)}),\ee
 where $\Pi_T^{(2,2)}$ is the five-dimensional gauge covariant version of the projector
\cite{graph}, \cite{KmcA} on the space of covariant analytic
transverse superfields $v^{++}.$ The properties of the
$\Pi^{(2,2)}_T(\zeta_1,\zeta_2)$ were described in the paper
\cite{K}.

For later, it is convenient to rewrite the projector
$\Pi^{(2,2)}_T(\zeta_1,\zeta_2)$ in a different form. Here we follow
the work \cite{KmcA}. We have \footnote{Here we are using  a  manifestly analytic form of
the delta function:
$$\d^{(2,2)}_A(1,2)=\f{1}{2\sB_1}(D^+_1)^4(D^+_2)^4\d^{13}(z_1-z_2)(D^{--}_2)^2\d^{(-2,2)}(u_1,u_2)$$}
\be
\Pi_T^{(2,2)}(\zeta_1,\zeta_2)=\d^{(2,2)}_A(1,2)-\Pi_L^{(2,2)}(1,2)\ee
$$=\d^{(2,2)}_A(1,2)-\cD^{++}_1\cD^{++}_2\f{1}{\sB_1}(D^+_1)^4(D^+_2)^4\d^{13}(z_1-z_2)\f{(u^-_1u^-_2)}{(u^+_1u^+_2)^3}~.$$
The  operator ${\sB}$ is defined by (\ref{covbox}). As a result,
$\Pi_T^{(2,2)}$ can be expressed as
\be\Pi_T^{(2,2)}(\zeta_1,\zeta_2)=\f{1}{\sB_1}[\cD^{++}_1,\sB_1]\f{1}{\sB_1}(D^+_1)^4(D^+_2)^4\d^{13}(z_1-z_2)\f{(u^-_1u^+_2)}{(u^+_1u^+_2)^3}\ee$$-\f{1}{\sB_1}(D^+_1)^4(D^+_2)^4\d^{13}(z_1-z_2)\f{1}{(u^+_1u^+_2)^2}.$$
Note that the first term does not vanish since
$[\cD^{++},\sB]\Phi^{(q)}=\f14(\cD^{+2}\cW)(1-q)\Phi^{(q)}$.

The next step is a representation of a two-point function in the
form
$$(\cD^+_1)^4(\cD^+_2)^4\f{1}{(u^+_1u^+_2)^q }=$$
\be\label{2point}=(\cD^+_1)^4\{(\cD_1^-)^4(u^+_1u^+_2)^{4-q}-\f14(u^+_1u^+_2)^{3-q}(u^-_1u^+_2)\Delta^{--}_1-(u^+_1u^+_2)^{2-q}(u^-_1u^+_2)^2\sB_1\}\ee
$$+\f14(q-3)(u^+_1u^+_2)^{1-q}(u^-_1u^+_2)^3(\cD^+\cD^+\cW),$$
where
\be\Delta^{--}=i\cD^{\a\b}\cD^-_\a\cD^-_\b+\cW(\cD^-)^2+4(\cD^{-\a}\cW)\cD^-_\a+(\cD^-\cD^-\cW).\ee
This representation was obtained in the  work \cite{khyper}. As a
result one gets the following expression for $\Pi^{(2,2)}_T$:
\be\label{pro}\Pi^{(2,2)}_T(1|2)=(u^-_1u^+_2)^2(\cD_1^+)^4\d^{13}(z_1-z_2)-\f{1}{\sB_1}\biggl\{(u^+_1u^+_2)-\f14(u^-_1u^+_2)(D^{+2}\cW)\f{1}{\sB_1}\biggr\}\ee
$$\times(\cD^+_1)^4\biggl\{(\cD^-_1)^4(u^+_1u^+_2)-\f14(u^-_1u^+_2)\Delta_1^{--}\biggr\}\d^{13}(z_1-z_2).$$
It is useful to compare this expression with a similar projector in
4D, $\cN=2$  SYM theory \cite{KmcA} .

\subsection{The deformed covariantly analytic d'Alembertian $\sB_\cW$}
The operator in the quadratic part of the action for quantum fields plays a
fundamental role in calculations of  the effective action in the
framework of the background field method. In the case under
consideration this operator is $\sB_{\cW}$ (\ref{sB}).

In the analytic subspace the operator $\sB_{\cW}$ (\ref{sB}) is
rewritten in the form
$$\f12(\cD^+)^4\cW(\cD^{--})^2=(-\f{1}{64})\biggl\{\cW(\cD^+)^4+
4(\cD^{+\hat\a}\cW)\cD^+_{\hat\a}(\cD^+)^2+((\cD^+)^2\cW)(\cD^+)^2\biggr\}(\cD^{--})^2.
$$
The first term here is, up to the multiplier $\cW$, the standard
covariant analytic d'Alembertian (\ref{covbox})
$$-\f{1}{64}\cW(\cD^+)^2(\cD^+)^2(\cD^{--})^2=\cW[\cD^{\hat{a}}\cD_{\hat{a}}-
\f14(\cD^{+\hat\a}\cD^+_{\hat\a}\cW)\cD^{--}+(\cD^{+\hat\a}\cW)\cD^-_{\hat\a}+
\f14(\cD^{-\hat\a} \cD^+_{\hat\a}\cW) -\cW^2]$$$$=\cW\sB.
$$
The deformation of the operator defined by (\ref{sB}) is stipulated
by the superfield $\cW$ as follows
$$(-\f{1}{64})((\cD^+)^2\cW)(\cD^+)^2(\cD^{--})^2=(-\f{1}{64})((\cD^+)^2\cW)[2(\cD^-)^2+16\cW\cD^{--}],$$
$$(-\f{1}{64})4(\cD^{+\hat\a}\cW)\times\cD^+_{\hat\a}(\cD^+)^2(\cD^{--})^2=(-\f{1}{64})4(\cD^{+\hat\a}\cW)\times[ 16(\cD^+_{\hat\a}\cW)\cD^{--}$$
$$-8\cW\cD^-_{\hat\a}-8i\cD_{\hat\a\hat\b}\cD^{-\hat\b}-16(\cD^-_{\hat\a}\cW)].$$
Summing up all together we obtain
\be\label{box}\sB_{\cW}=
\cW\bigg\{\cD^{\hat{a}}\cD_{\hat{a}}-[\f{D^{+\hat\a}\cW
D^{+}_{\hat\a}
\cW}{\cW}+\f12(\cD^{+\hat\a}\cD^+_{\hat\a}\cW)]\cD^{--}-\f{1}{32}\f{(D^{+2}\cW)}{\cW}(\cD^-)^2\ee
$$+\f32(\cD^{+\hat\a}\cW)\cD^-_{\hat\a}+\f{1}{2\cW}(D^{+\hat\a}\cW)i\cD_{\hat\a\hat\b}\cD^{-\hat\b}+
\f{D^{+\hat\a}\cW D^-_{\hat\a}\cW}{\cW}+\f14\cD^{-\hat\a}
\cD^+_{\hat\a}\cW -\cW^2\bigg\}.$$
 This is a final result for the operator $\sB_{\cW}$ acting on analytic superfields.

The deformed covariantly analytic d'Alemberian $\sB_{\cW}$ possesses
a useful property $[\cD^+_{\hat\a},\sB_{\cW}]=0 .$ It is important
to note that the coefficient at the harmonic derivative $\cD^{--}$ in
(\ref{box}) is $-\f{1}{\cW}{\cal G}^{++}$ and that on the equations
of motion (\ref{EoM}) this term vanishes! Also on the equations of
motion $D^{--}{\cal G}^{++}=0$ and therefore
$$\f{D^{+\hat\a}\cW D^-_{\hat\a}\cW}{\cW}+\f14(D^-D^+\cW)=-\f14(D^+
D^-\cW).$$ This means that in this case the operator $\sB_{\cW}$ takes
the form
\be\label{boxonshell} \sB_{\cW}=\cW\bigg\{\cD^{\hat{a}}\cD_{\hat{a}}-\f{1}{32}\f{(D^{+2}\cW)}{\cW}(\cD^-)^2+\f32(\cD^{+\hat\a}\cW)\cD^-_{\hat\a}+\f{1}{2\cW}(D^{+\hat\a}\cW)i\cD_{\hat\a\hat\b}\cD^{-\hat\b}\ee
$$
-\f14(D^+
D^-\cW) -\cW^2\bigg\}.$$

\subsection{ The leading contribution to effective action of the 5D SCS multiplet}

Our next aim is to calculate the leading low-energy quantum
contribution to the one-loop effective action. To do that it is
sufficient to evaluate the one-loop effective action
$\G^{(1)}_{SCS}[V^{++}]$ for an on-shell background vector
multiplet. The representation (\ref{propt}) provides a simple and
powerful scheme for computing the effective action in the framework
of 5D, $\cN = 1$ superfield proper-time technique\footnote{ Note that
$\Tr \ln \cW$ does not contribute to (\ref{oneloop}).}.

If the background gauge multiplet is on-shell, ${\cal G^{++}} = 0$,
the analytic d’Alembertian $\sB_\cW$ does not involve any harmonic
derivative $\cD^{--}$, but $\sB$ contains the factor
$\cD^{+2}\cW\cD^{--}$. Hence, we get
$$\f{1}{\sB_1}(u^+_1u^+_2)=(u^+_1u^+_2)\f{1}{\sB_1}+
\f{1}{\sB_1}\f14(D^{+2}\cW)(u^-_1u^+_2)\f{1}{\sB_1}.$$ Taking into
account this relation we obtain that the second term in the operator
$\Pi^{(2,2)}$ is absent. Therefore the projector is reduced only to
$(\cD^+)^4\d^{13}(z_1-z_2).$ Then
\be\label{props}\G^{(1)}_{SCS}[V^{++}]=-\f{i}{2}\int\f{ds}{s}\int
d\zeta^{(-4)} e^{is\sB_\cW}(\cD^+)^4\d^{13}(z_1-z_2)|_{z_1=z_2}.\ee
Now we replace the delta-function in (\ref{props}) by its Fourier
representation \be\d^{5}(x_1-x_2)(\cD_1^+)^4\d^8(\t_1-\t_2)=\int
\f{d^5p}{(2\pi)^5}e^{ip_{\hat{a}}\rho^{\hat{a}}}\d^4(\t^+_1-\t^+_2),\ee
where
$\rho^{\hat{a}}=(x_1-x_2)^{\hat{a}}-2i(\t^+_1-\t^+_2)\G^{\hat{a}}\t^-_2$
is the supersymmetric interval. Then we push
$e^{ip_{\hat{a}}}\rho^{\hat{a}}$
 through all the operator factors in (\ref{props}) to the left and then
replace it by unity in the coincidence limit. This leads to the
following transformation of the covariant derivatives
\be\cD_{\hat{a}}\rightarrow \cD_{\hat{a}}+ip_{\hat{a}}, \q
\cD^-_{\hat\a}\rightarrow \cD^-_{\hat\a}
-2ip_{\hat\a\hat\b}(\t_1-\t_2)^{-\hat\b}.\ee Note that  the shift
$-2ip_{\hat\a\hat\b}(\t_1-\t_2)^{-\hat\b}$  in $\cD^-_{\hat\a}$
vanishes in the coincidence limit since there is no enough
operators $\cD^+_{\hat\a}$ to annihilate
$(\t^-_1-\t^-_2)$. To get the expansion of the effective action in the
background fields and their derivatives we should expand  the
exponent of the  operator and calculate the standard momentum
integrals. Thus, we expand the exponent $e^{is\sB_{\cW}}$ in powers
of spinor derivatives near $e^{-isp^2}$ up to the fourth order in
spinor derivatives (it corresponds to expanding up to the fourth
order in the proper time) and use the identity
$-\f12(\cD_1^-)^4\d^4(\t^+_1-\t^+_2)|_{\t_1=\t_2}=1$. It remains to
perform a standard integration over the momentum variables and over the
proper-time. The final result has the form
\be\label{impro}\G^{(1)}_{SCS}[V^{++}]=\f{5}{(16\pi)^2}\int
d^{13}z\cW\ln\f{\cW}{\Lambda}~.\ee
 Here $\Lambda$ is some scale\footnote{It is easy to see that the effective action (\ref{impro})
 does not depend on the scale, since $$\int d^{13}z\cW\ln\Lambda=\ln\Lambda\int d\zeta^{(-4)}(D^+)^4\cW=0.$$}.
One can show, using the methods developed in \cite{Kall}, that the
action (\ref{impro}) is superconformal. A component form of
(\ref{impro}) in the bosonic sector corresponds to the Lagrangian
$\f{F^4}{{\phi}^3}$.

The effective action (\ref{impro}) can be treated as the 5D analogue
of the non-holomorphic potential in $4D, \cN=2$ supersymmetric gauge
theories. This effective action is also similar to the action of the
the so-called 4D improved tensor multiplet \cite{WR82}.

\section{The leading contribution to the effective action of a 5D SYM multiplet}
It is interesting and instructive to compare the one-loop leading
low-energy effective action in  5D SCS theory and in 5D SYM
theory.

5D, $\cN=1$ SYM theory is described by the action (see (\ref{V^--}),
(\ref{SYM})): \be\label{SYM-1} S_{SYM}=\f{1}{g^2_{SYM}}\int d^{13}z
\ V^{++}V^{--}~,\ee
where the coupling constant has
dimension $[g^2_{SYM}]=1.$

Let us consider the construction of the background field method for the theory
(\ref{SYM-1}). It is obvious that the second variation of the action
has the form completely analogous to the 4D, $\cN=2$ case
\be\label{delta2}\d^2 S=\f{1}{g^2_{SYM}}\int d^{13}z du_1du_2 \d
V^{++}(u_1)\d V^{++}(u_2)\f{1}{(u^+_1u^+_2)^2}.\ee
 Therefore we can simply
repeat step by step all the stages of the construction of  the effective
action developed in \cite{backgr}, \cite{KmcA}. In particular, we can
use the same choice of the gauge  conditions on the quantum
superfield $D^{++}v^{++}=0$ and the same procedure to fix  gauge as
in \cite{backgr}. It gives the sum of the quadratic part of action
(\ref{delta2}) and the corresponding gauge-fixed action in the form
 $S_{GF}$
$$S_2+S_{gh}=\f12(1+\f{1}{\a})\int d^{13}z du_1du_2\f{1}{g^2_{SYM}}\f{v^{++}(1)v^{++}(2)}{(u^+_1u^+_2)^2}$$
$$-\f{1}{2\a}\int d^{13}z du\f{1}{g^2_{SYM}}v^{++}\f12(D^{--})^2v^{++}$$
Further we put gauge parameter $\a=-1$.

The derivation of the general expression for the one-loop effective action
in 5D SYM theory is completely analogous to derivation of the
expression (\ref{oneloop}). The final result is
\be\label{oneloopsym}
\tilde{\Gamma}^{(1)}_{SYM}=-\f{i}{2}\Tr\ln(\cD^{++})^2+\f{i}{2}(\Tr_{(2,2)}\ln
\sB -\Tr_{(0,4)}\ln \sB). \ee
 The first term is the contribution of
the ghosts and the second term is the contribution of a SYM multiplet.
Here the covariantly analytic d'Alembertian $\sB$ is given by
 (\ref{covbox}) and has the form \cite{khyper}
\be\label{covanB}\sB=-\f{1}{64}(\cD^+)^4(\cD^{--})^2|=
\cD^{\hat{a}}\cD_{\hat{a}}+(\cD^{+\hat\a}\cW)\cD^-_{\hat\a}-\f14(\cD^{+\hat\a}\cD^+_{\hat\a}\cW)\cD^{--}\ee$$
+\f14(\cD^{-\hat\a} \cD^+_{\hat\a}\cW) -\cW^2 .
$$

Now we consider a calculation of the contribution to the
effective action (\ref{oneloopsym}) from a SYM multiplet. Contribution
of the first term in (\ref{oneloopsym}) will be calculated in
section 5. In the proper-time representation, the  contribution of
SYM multiplet has the form analogous to (\ref{propt})
\be\label{propt-1}\G^{(1)}_{SYM}=-\f{i}{2}\int_0^\infty
\f{ds}{s}\Tr(e^{is\sB}\Pi_T^{(2,2)}).\ee The only difference with
(\ref{propt}) is the operator $\sB$ instead of the operator
$\sB_{\cW}$.
 To find the leading low-energy contribution to effective action it is sufficient to
 consider the on-shell background superfield. Classical on-shell equation for the
5D SYM theory look like $(\cD^+)^2\cW=0$. In this case the
covariantly  analytic d'Alembertian $\sB$ and the projection
operator $\Pi^{(2,2)}_T$ are simplified and we have
the effective action in the following form
$$\Gamma^{(1)}_{SYM}=-\f{i}{2} \int_0^\infty\f{ds}{s}\Tr e^{is\sB} \Pi^{(2,2)}_T=
-\f{i}{2}\int_0^\infty\f{ds}{s}\int d\zeta^{(-4)}
e^{is\sB}(\cD^+)^4\d^{13}(z_1-z_2),$$
 Evaluation of this expression is realized completely the same way as it was
 done in the subsection 3.3 to obtain the (\ref{impro}). The final
 result has the form
\be\label{impro-1}\Gamma^{(1)}_{SYM}=\f{1}{24\pi^2} \int d^{13}z \
\cW \ln \f{\cW}{\Lambda}.\ee
 We see that the functional form of the effective action (\ref{impro-1}) generated by a SYM multiplet
 coincides with the one generated by a SCS multiplet (\ref{impro}).
 Although the on-shell conditions for 5D SCS theory and for 5D SYM theory
 are different and, moreover,  5D SYM theory is characterized by a
 dimensional coupling constant, the leading low-energy contributions
in these two theories to the one-loop effective action turn out to be the same up to a numerical
 coefficient.

\section{ The leading and next-to-leading contributions of the ghosts and matter superfields}
The ghost contribution to the one-loop effective action in both SCS  and
SYM theories  is defined by the expression
\be\label{ghost}\G^{(1)}_{ghost}= -\f{i}{2}\Tr\ln (\cD^{++})^2. \ee
The contribution from matter hypermultiplet superfields differs only by a
sign and by the choice of the representation of the gauge group. Hence, to find the
leading contribution to (\ref{ghost}) we can use the results from
the previous analysis \cite{khyper} of the effective action for a
$q$-hypermultiplet coupled to a background vector
multiplet\footnote{ From a formal point of view,  the   effective action (\ref{ghost})
corresponds to a so called $\omega$-hypermultiplet
\cite{gios}. It was pointed out some time ago \cite{backgr} that
this effective action is equal, up to the coefficient $2,$ to the
effective action for a  $q$ hypermultiplet. Therefore, further we will
take into account just the $q$ hypermultiplet effective action}. It
was shown in that paper that the leading quantum correction is the
SCS action. Therefore, further we will focus only on
the first next-to-leading correction.

For calculation of the first next-to-leading contribution to the
effective action (\ref{ghost}) we will follow the procedure proposed
in the papers \cite{KmcA}, \cite{khyper}. This procedure is based on
calculating the variation of the effective action with its subsequent
restoration  given the obtained variation.

Let  $\G^{(1)}_{hyper}$ be the one-loop contribution to the effective
action from either matter hypermultiplets or  from ghosts
 \be\label{eahyp} \G^{(1)}_{hyper} = (\pm)i\Tr \ln \cD^{++}=\mp i\Tr\ln G^{(1,1)}~,
 \ee
 where the upper sign corresponds to the contribution from matter
and the lower sign to that of ghosts. Further, for simplicity, we will consider for
 only the plus sign. Here $G^{(1,1)}(\zeta_1,\zeta_2)$ is the
hypermultiplet Green function satisfying the equation
\be\label{Greenhyper}\cD^{++}_1G^{(1,1)}(\zeta_1,\zeta_2)=\d_A^{(3,1)}(\zeta_1,\zeta_2)~,\ee
where $\d_A^{(3,1)}(\zeta_1,\zeta_2)$ is the appropriate covariantly
analytic delta-function
$$\d_A^{(3,1)}(\zeta_1,\zeta_2)=(\cD_1^+)^4\d^{13}(z_1-z_2)\d^{(-1,1)}(u_1,u_2){\bf 1}.$$
This equation is similar to the equation for the hypermultiplet Green
function in 4D, $\cN=2$ theories and we can use  methods
developed in \cite{backgr}. The
Green function $G^{(1,1)}(\zeta_1,\zeta_2)$ can be written in the
form \be\label{prophyper}G^{(1,1)}(\zeta_1,\zeta_2)=
-\f{1}{\sB_1}(\cD^+_1)^4(\cD^+_2)^4\d^{13}(z_1-z_2)\f{1}{(u^+_1u^+_2)^3}{\bf
1}.\ee Here $\sB$ is the covariantly analytic d'Alembertian
(\ref{covanB}) and $(u^+_1u^+_2)^{-3}$ is a special harmonic
distribution \cite{graph}.

The variation of the effective action (\ref{eahyp}) under the
background superfield $V^{++}$ is written as follows (see
\cite{KmcA}, \cite{khyper} for details)
\be\label{varG}\d\G^{[1]}_{hyper}=-\int d\zeta^{(-4)}\bigg\{\d
V^{++}G^{(1,1)}\bigg\}\ee
 To evaluate this expression one can use the proper-time technique.
The leading low-energy contribution goes from the terms without
derivatives of $\cW$. It was calculated in \cite{khyper} and has the
form \be\label{varCS}\d\G^{[1]}_{hyper}=-\f{1}{(4\pi)^2}{\rm
sign}(\cW)\int d^{13}zdu  \d V^{++}V^{--}\cW~.\ee The variation of
(\ref{varCS}) corresponds precisely to the classical action
(\ref{CSaction}) of the SCS gauge theory.

Our main purpose in this section is to find the first
next-to-leading correction to the SCS action. First of all one considers
the Dyson type equation relating the free and full hypermultiplet
propagators \cite{backgr}
\be\label{GG0}G^{(1,1)}(1|2)=G_0^{(1,1)}(1|2)-\int d\zeta^{(-4)}_3
G_0^{(1,1)}(1|3)i V^{++}(3)G^{(1,1)}(3|2)~.\ee Substituting
eq.(\ref{GG0}) into the variation (\ref{varG}), one finds
$$\d\G^{(1)}_{hyper}=\int d\zeta^{(-4)}_1 d\zeta^{(-4)}_2\d V^{++}(1)iV^{++}(2)G_0^{(1,1)}(1|2)G^{(1,1)}(2|1)~.$$
Taking into account the explicit form of the propagator
(\ref{prophyper}),  we rewrite this expression as follows
\be\label{exact}\d\G^{(1)}_{hyper}=\int d\zeta^{(-4)}_1
d\zeta^{(-4)}_2\d
V^{++}(1)\f{1}{\Box_1}(D^+_1)^4(D^+_2)^4\f{\d^{13}(z_1-z_2)}{(u^+_1u^+_2)^3}\ee$$\times
iV^{++}(2)\f{1}{\sB_2}(\cD^+_2)^4(\cD^+_1)^4\f{\d^{13}(z_2-z_1)}{(u^+_2u^+_1)^3}~.$$
Now we  use the spinor derivatives from the first delta-function to
restore the full $\cN =1$ superspace measure according to the rule
$\int d\zeta^{(-4)}(D^+)^4=\int d^{13}z$. So far, we did not
consider any restrictions for the background superfield, therefore
(\ref{exact}) is the exact representation for the one-loop
hypermultiplet effective action. In principle it can be a starting
point for calculations of different contributions to the one-loop
effective action.

To compute the first next-to-leading quantum corrections to
(\ref{varCS}) it is sufficient to expand in (\ref{exact}) the
covariant analytic d'Alembertian in powers  of the derivatives of
the background superfields. Let us remember that the Green function
of the hypermultiplet is antisymmetric with respect to the permutation
of its arguments. This allows us to rewrite the expression
(\ref{exact}) in the form \be\label{exact1} \d\G^{[1]}_{hyper}=-\int
d^{13}z du_1du_2\f{\d
V^{++}(1)}{(u^+_1u^+_2)^3}\f{1}{\Box}(\cD^+_1)^2iV^{++}(2)\f{1}{\sB_1}(-\f{1}{32})(\cD^+_1)^2(\cD^+_2)^4\f{\d^{13}(z_2-z_1)}{(u^+_1u^+_2)^3}.\ee
The next step is the expansion of the operator $\sB$ in the denominator
in power series of the derivatives of $\cW$.  In this expansion we keep only  the leading
terms with two derivatives. It leads to
$$\d\G^{(1)}_{hyper}=-\int d^{13}z du_1du_2 \f{\d V^{++}(1)}{(u^+_1u^+_2)^2}(\cD^-_2)^2iV^{++}(2)\f{1}{\Box}\f{\cD_1^+\cW\cD_1^+\cW}{(\Box-\cW^2)^3}(-\f14)(\cD^-_2)^4(\cD^+_2)^4\d^{13}(z_2-z_1).$$
Now the usual steps lead to
$$\d\G^{[1]}_{hyper}= \f{i}{6(4\pi)^2}\int d^{13}z du\d V^{--}\f{\cD^+\cW\cD^+\cW}{\cW^2}$$$$=
-\f{i}{6(4\pi)^2}\int d^{13}z  du \d V^{--}  (\cD^+)^2\ln\cW=
-\f{1}{12\pi^2}\int d^{13}z   \d \cW \ln\cW~.$$
 As a result we obtain the first next-to-leading contribution to the
 one-loop hypermultiplet effective action in the form
\be\label{hyperresult}\G^{[1]}_{hyper}=c_{hyper} \int d^{13}z \ \cW\
\ln\f{\cW}{\Lambda}~.\ee
 Here $c_{hyper}$ is a numerical coefficient depending on details of the
 hypermultiplet action (such as the number of components, the representation, whether hypermultiplet superfields are commuting or anticommuting).
 We see that the effective action (\ref{hyperresult}) has the same functional
structure as the effective actions generated by the SCS multiplet
(\ref{impro}) and by the SYM multiplet (\ref{impro-1}).

\section{Summary}
We have considered the five-dimensional $\cN=1$ supersymmetric field
models such as the Abelian Chern-Simons theory, the Yang-Mills
theory and the hypermultiplet theory coupled to a background vector
multiplet, formulated in the harmonic superspace approach. In all
these models we calculated the universal four derivative
contributions to the one-loop effective action.

In the 5D SCS theory we have developed the background field method
in harmonic superspace and represented the one-loop effective action
in terms of functional determinants of the operators acting in the
analytic subspace of harmonic superspace (Eq. (\ref{oneloop})). The
above expression contains the contributions from the ghost
superfields and from a SCS vector multiplet superfield. We studied the
structure of the latter contribution to the effective action and
evaluated the leading low-energy effective action (Eq.
(\ref{impro})).

The same consideration has been realized in 5D SYM theory as well.
We developed the background field method for this theory, found the
one-loop effective action in terms of the functional determinants of
the differential operators acting in the analytic subspace of
harmonic superspace (Eq. (\ref{oneloopsym})). We studied the structure
of the one-loop contributions to the effective action from a SYM vector
multiplet and calculated the leading low-energy contribution of this
multiplet. Although  5D SYM theory is not superconformal, its
coupling constant is dimensional, and  the expressions
(\ref{oneloop}) and (\ref{oneloopsym}) are different, the leading
contribution to the effective action in  5D SYM theory (Eq.
(\ref{impro-1})) has the same functional form as in  5D SCS
theory.

In  5D hypermultiplet theory in a vector multiplet background
field we have calculated the first next-to leading contribution to the
effective action (Eq. (\ref{hyperresult})). The corresponding
leading contribution is the SCS action and was found in the paper
\cite{khyper}. We have shown that this first next-to leading
contribution (\ref{hyperresult}) again has the same functional form
as the leading contribution in SCS theory.

We have found the manifestly 5D, $\cN = 1$ superconformal form of
the term $\sim F^4$ in the effective actions of  the SCS theory,
 SYM theory and   hypermultiplet theory. The next step of
studying of the one-loop effective action in the theories under
consideration is a construction of the full low-energy one-loop
effective actions where  all the powers of the Abelian strength are
summed up. In other words, the next purpose is to construct the 5D
superfield Heisenberg-Euler type of the effective action. Like in
4D, 3D superconformal gauge theories \cite{BKT-BPS} it is reasonable
to expect that this effective action will be expressed in the terms
of so called superconformal invariants which transform as
scalars under the 5D, $\cN = 1$ superconformal group. For example,
such a scalar invariant can be a superfield
$\Psi^2=\f{1}{\cW^2}D^4\ln\cW.$ We hope to study this issue in the
forthcoming work.

\section*{Acknowledgments }

The authors are very grateful  to E.A. Ivanov and S.M. Kuzenko for useful comments.
The authors are thankful to the RFBR grant, project No 15-02-06670-а
and LRSS grant, project No 88.2014.2 for partial support. Work of
I.L.B was supported by Ministry of Education and Science of Russian
Federation, project No 2014/387/122. Also I.L.B is grateful to the
DFG grant, project LE 838/12-2.


\bigskip

\end{document}